\documentclass{appolb}
\usepackage{epsfig}
\usepackage{wrapfig}

\begin{document}
\pagestyle{plain}
\newcount\eLiNe\eLiNe=\inputlineno\advance\eLiNe by -1
\title{REPORT FROM NA49 
}

\author{Katarzyna GREBIESZKOW  \\{\small for the NA49 Collaboration}
\address{Faculty of Physics, Warsaw University of Technology,
Koszykowa 75, 00-662~Warsaw, Poland}}
\maketitle

\begin{abstract}

The signatures of the onset of deconfinement, found by the NA49 
experiment at low SPS energies, are confronted with new results from 
the Beam Energy Scan (BES) program at BNL RHIC and CERN LHC 
results. Additionally, new NA49 results on chemical (particle ratio) 
fluctuations, azimuthal angle fluctuations, intermittency of di-pions, 
etc. are presented. 

\end{abstract}

\section{Introduction}

The NA49 experiment \cite{na49_nim} at the CERN SPS, taking 
data in 1994-2002, studies an important region of the phase diagram of 
strongly interacting matter. First, the energy threshold for 
deconfinement (minimum energy to create a partonic system) was found 
at low SPS energies \cite{mg_model, na49_kpi}. Second, theoretical 
QCD-based calculations suggest that the critical point of strongly 
interacting matter is located at energies accessible at the CERN SPS 
accelerator (i.e. $T^{CP} = 162 \pm 2$ MeV, $\mu_B^{CP} = 360 \pm 40 $ 
MeV) \cite{fodor_latt_2004}).

\section{Onset of deconfinement}

The NA49 energy scan program (completed in 2002) was motivated by 
predictions of the Statistical Model of the Early Stage (SMES) 
\cite{mg_model} assuming that the energy threshold for deconfinement 
is located at low SPS energies. Several structures in excitation functions 
were expected within the SMES: a kink in the increase of the pion yield per
participant nucleon (change of slope due to increased entropy as a 
consequence of the activation of partonic degrees of freedom), a sharp 
peak (horn) in the strangeness to entropy ratio, and a step in the 
inverse slope parameter of transverse mass spectra (constant 
temperature and pressure in a mixed phase). Such signatures were indeed 
observed in $A+A$ collisions by the NA49 experiment \cite{na49_kpi}, 
thus locating the onset of deconfinement (OD) energy around 30$A$ GeV 
($\sqrt{s_{NN}} \approx 7.6$~GeV).

\subsection{Verification of NA49 results and interpretation by STAR and 
ALICE}

\begin{wrapfigure}{r}{6.cm}
\vspace{-0.5cm}
\includegraphics[scale=0.31]{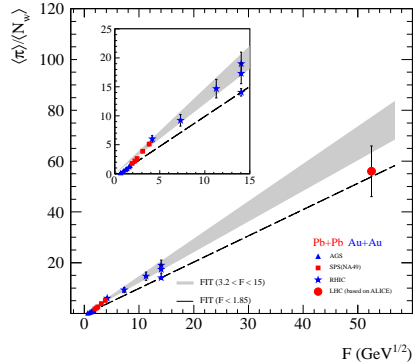}
\vspace{-1.1cm}
\caption[]{\footnotesize Mean pion multiplicity per participant 
nucleon. See \cite{anar} for details.}
\label{kink}
\vspace{-0.5cm}
\end{wrapfigure}

Until recently the evidence of OD was based on the results of a single 
experiment. Recently new results on central Pb+Pb collisions at the LHC 
\cite{lhc} and data on central Au+Au collisions from the RHIC BES 
program \cite{bes} were released. Figure \ref{kink} shows an update 
of the kink plot, where BES points follow the line for $A+A$ 
collisions and the LHC point~\footnote{the mean pion multiplicity at LHC 
was estimated based on the ALICE measurement of charged particle 
multiplicity, see \cite{anar} for details.}, within a large error, does 
not contradict extrapolations from high SPS and RHIC energies.

\begin{figure}[ht]
\centering
\vspace{-0.3cm}
\includegraphics[width=0.49\textwidth]{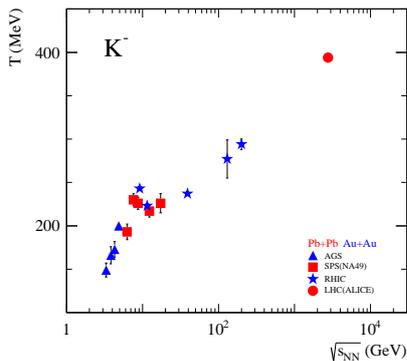}
\includegraphics[width=0.49\textwidth]{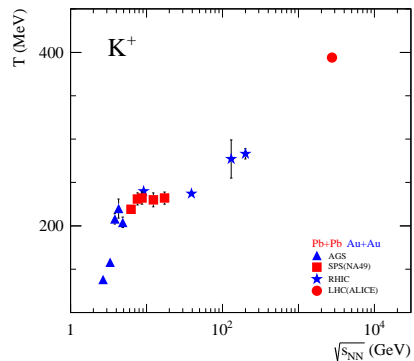}
\vspace{-0.7cm}
\caption[]{\footnotesize Inverse slope parameters of kaon $m_T$ 
spectra. See \cite{anar} for details.}
\label{step}
\end{figure}

\begin{wrapfigure}{r}{6.cm}
\vspace{-0.5cm}
\includegraphics[scale=0.31]{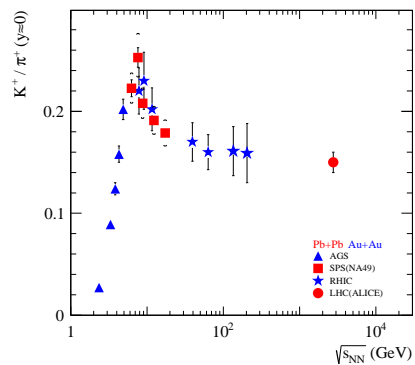}
\vspace{-1.1cm}
\caption[]{\footnotesize Kaon to pion yield (near midrapidity). See
\cite{anar} for details.}
\label{horn}
\vspace{-0.5cm}
\end{wrapfigure}

Figure \ref{step} shows inverse slope parameters of kaon transverse 
mass spectra. The LHC points and the RHIC BES points confirm the step 
structure expected for the onset of deconfinement. The $K^{+}/\pi^{+}$ yield 
(near midrapidity) is presented in Fig. \ref{horn}. As seen, 
RHIC results confirm NA49 measurements at the onset of 
deconfinement. Moreover, LHC (ALICE) data demonstrate that the energy 
dependence of hadron production properties shows rapid changes only at 
low SPS energies, and a smooth evolution is observed between the top 
SPS (17.2 GeV) and the current LHC (2.76 TeV) energies. 
All three structures confirm that results agree with the interpretation 
of the NA49 structures as due to OD. Above the onset energy only a 
smooth change of QGP properties with increasing energy is expected.

\section{New NA49 results on fluctuations}

Fluctuations and correlations may serve as a signature of 
the onset of deconfinement. Close to the phase transition the Equation of 
State changes rapidly which can impact the energy dependence of 
fluctuations. Moreover, fluctuations and correlations can help to 
locate the critical point (CP) of strongly interacting matter. This is 
in analogy to critical opalescence, where we expect enlarged 
fluctuations close to the CP. For strongly interacting matter a maximum of 
fluctuations is expected when freeze-out happens near the CP. Therefore the CP 
should be searched above the onset of deconfinement energy,
found by NA49 to be 30$A$ GeV ($\sqrt{s_{NN}} \approx 7.6$~GeV).

\subsection{Particle ratio fluctuations}

NA49 used $\sigma_{dyn}$ to measure dynamical particle 
ratio fluctuations. $\sigma_{dyn}$ is defined as the difference between the 
relative widths of particle ratio distributions for data and
for artificially produced mixed events, where only statistical
fluctuations are present (see \cite{tim_qm11} for details).
  
\begin{figure}[ht]
\centering
\vspace{-0.5cm}
\includegraphics[width=0.45\textwidth]{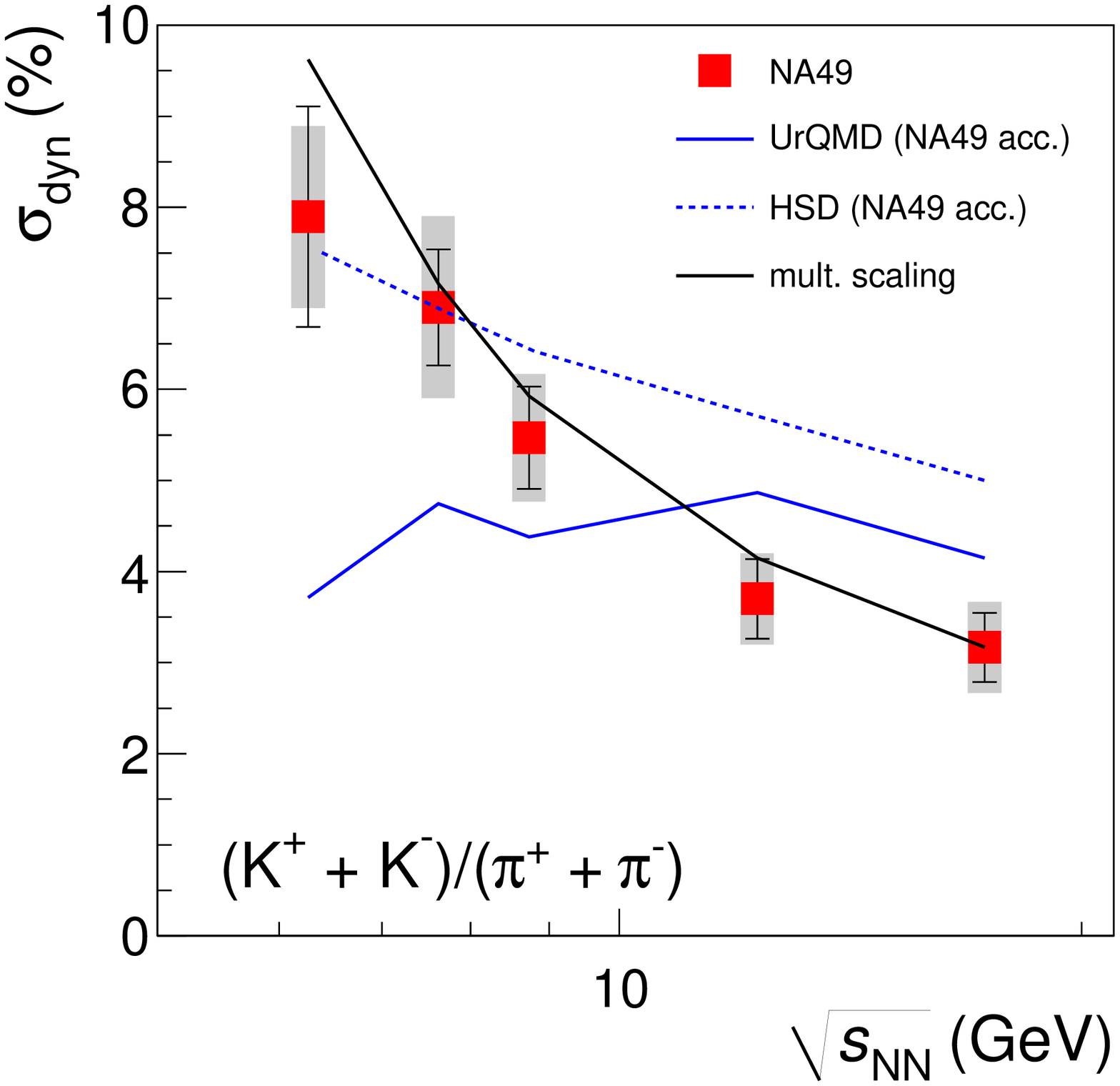}
\includegraphics[width=0.45\textwidth]{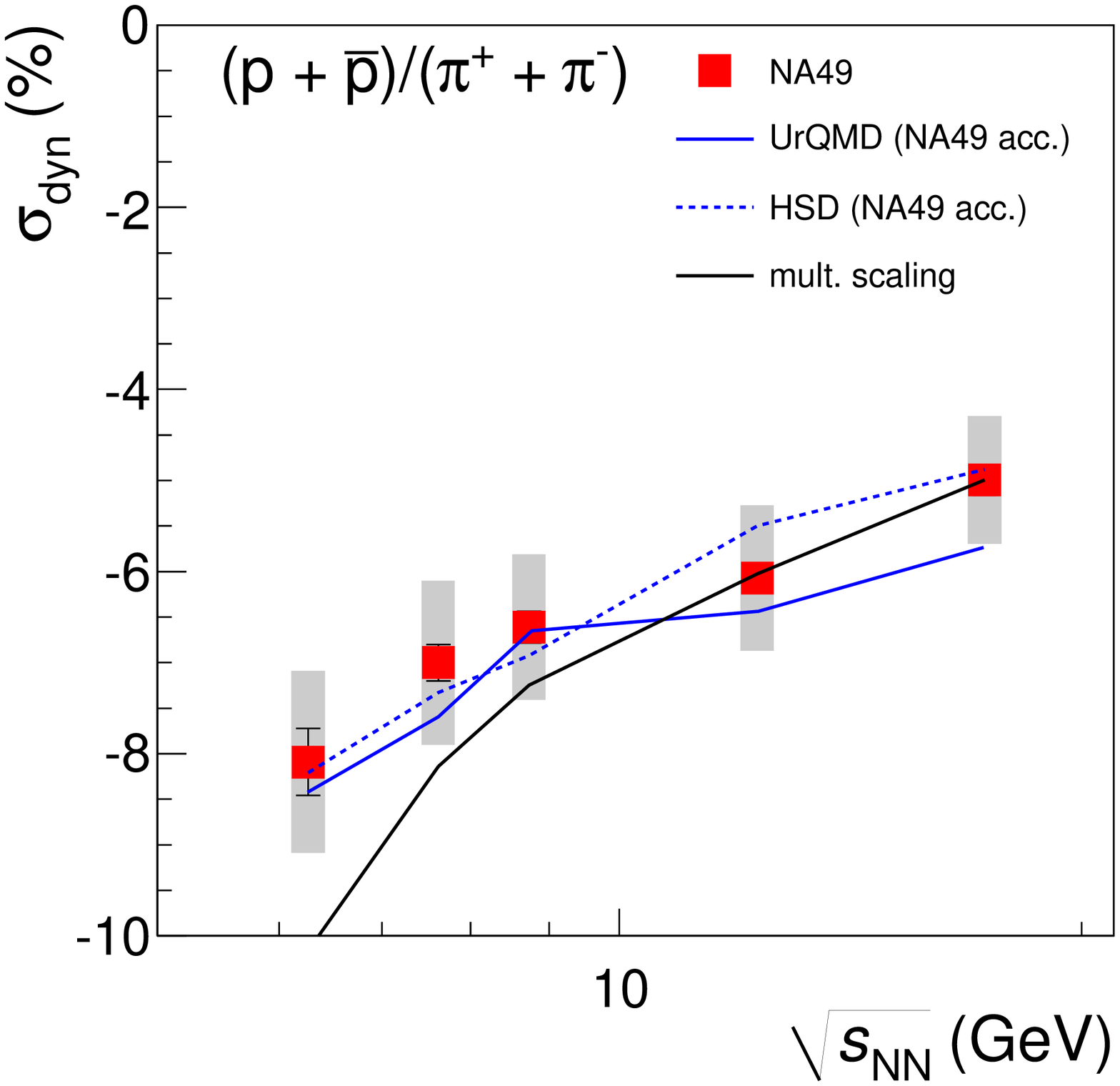}
\vspace{-0.3cm}
\caption[]{\footnotesize Energy dependence of $K/\pi$ and  
$p/\pi$ fluctuations \cite{tim_qm11}.}
\label{ratio1}
\end{figure}

The energy dependence of event-by-event fluctuations of the particle ratios 
$K/\pi$ and $p/\pi$ (for the 3.5\% most central Pb+Pb collisions) is shown in 
Fig.~\ref{ratio1}. $K/\pi$ fluctuations show positive values of $\sigma_{dyn}$. 
The steep rise towards low SPS energies is not reproduced by the UrQMD model. 
The HSD model catches the trend but over-predicts high energy SPS 
results. The $p/\pi$ ratio shows negative dynamical fluctuations. This 
behavior is reproduced by hadronic models and understood in terms of 
correlations due to nucleon resonance decays.

\begin{figure}[ht]
\centering
\vspace{-0.2cm}
\includegraphics[width=0.45\textwidth]{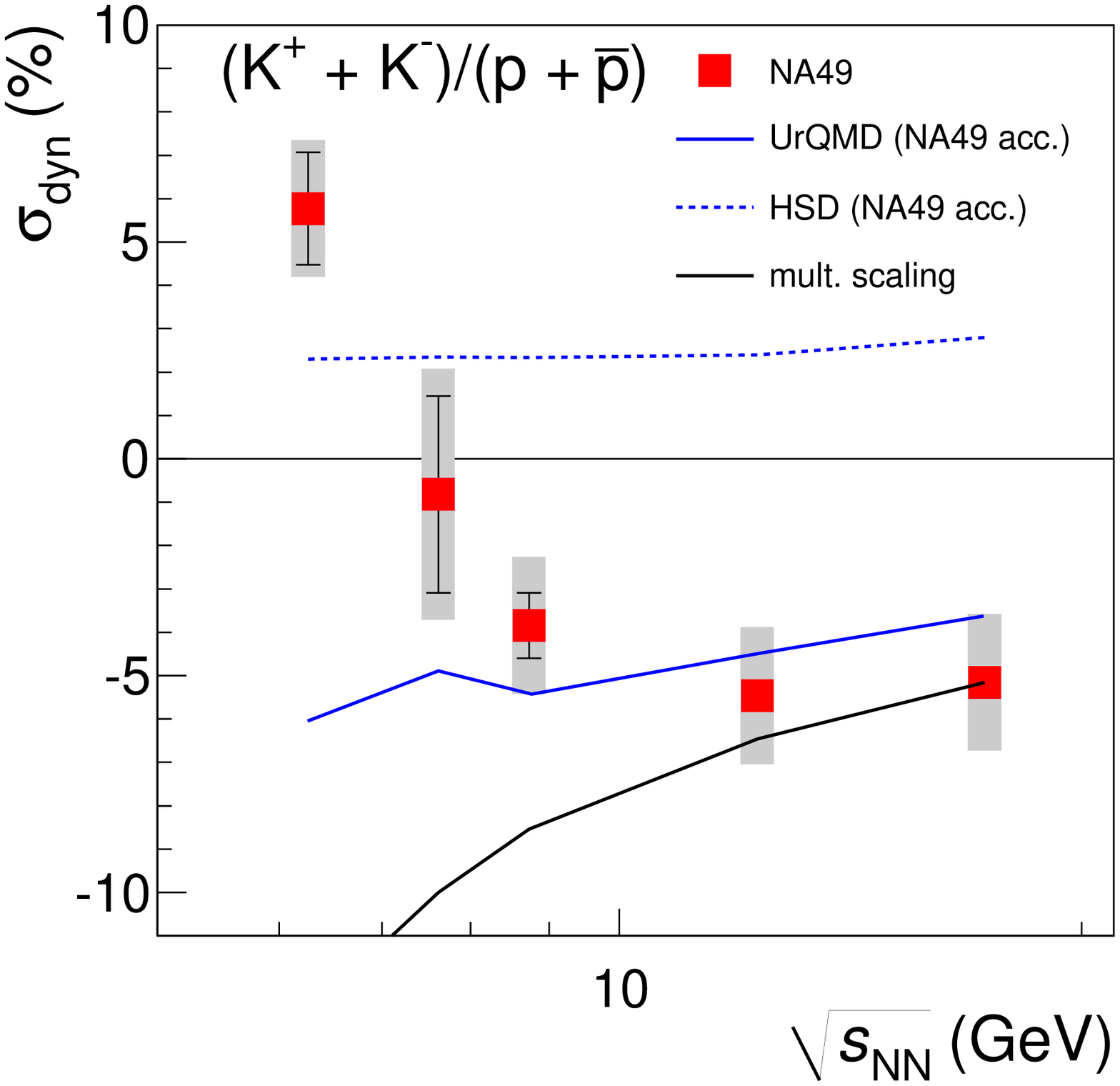}
\includegraphics[width=0.45\textwidth]{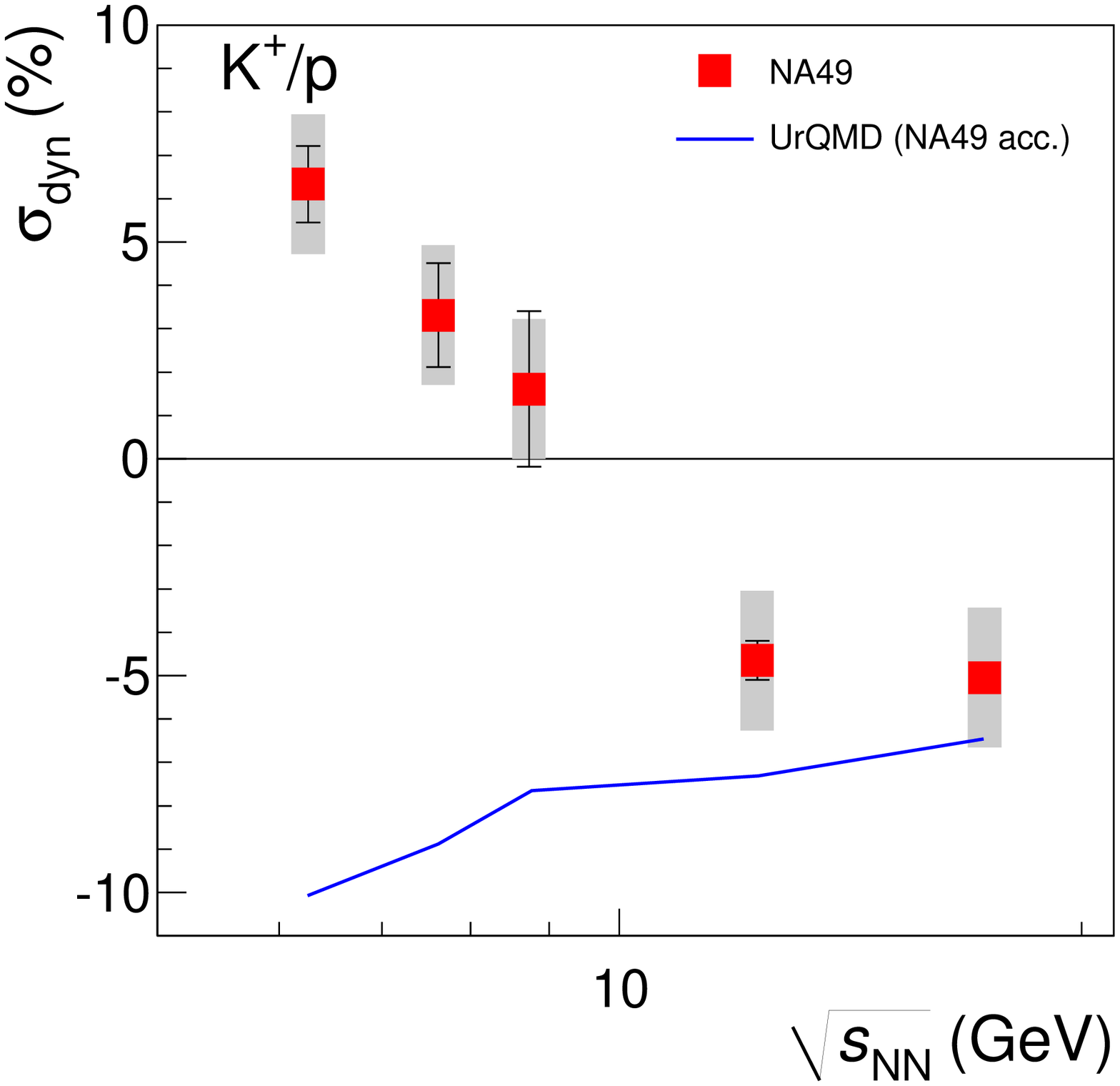}
\vspace{-0.3cm}
\caption[]{\footnotesize Energy dependence of $K/p$ fluctuations 
\cite{tim_qm11}.}
\label{ratio2}
\end{figure}

An unexpected result was obtained for event-by-event $K/p$ 
fluctuations (Fig. \ref{ratio2}). Dynamical fluctuations change sign 
close to the onset of deconfinement energy. A jump to positive values 
at lowest SPS energies is followed by a negative plateau at higher 
SPS energies. Such structure is not described by hadronic models (UrQMD 
and HSD). Additionally we show $K^{+}/p$ fluctuations in which no 
contributions from resonance production are expected. The relation of 
this intriguing result to the onset of deconfinement is not known yet.

It has been suggested \cite{tim} that $\sigma_{dyn}$ can be separated 
into two terms: a correlation strength term and a term purely dependent on 
multiplicities. In case of unchanged correlations (invariant correlation 
strength) the general expectation is
$\sigma_{dyn} \propto \sqrt{\frac{1}{\langle A \rangle}  + 
\frac{1}{\langle B \rangle}}$, 
where $A, B = N_{K}, N_{\pi}, N_{p}$, etc.
Such scaling is presented in Figs. \ref{ratio1}, \ref{ratio2} as black 
solid lines. One can see that scaling works very well for $K/\pi$ and 
$p/\pi$ fluctuations. The change of sign in $K/p$ fluctuations excludes 
any simple scaling based on average multiplicities. The above scaling 
assumed invariant correlation strength, therefore the NA49 results suggest 
that the underlying correlation between kaons and protons is changing 
with energy.

\begin{figure}[ht]
\centering
\vspace{-0.4cm}
\includegraphics[width=1.0\textwidth]{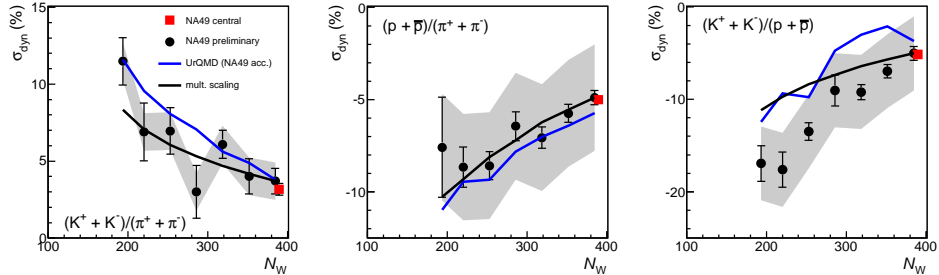}
\vspace{-0.8cm}
\caption[]{Centrality dependence of particle ratio fluctuations at 
158$A$ GeV \cite{tim_qm11}.}
\label{ratio3}
\end{figure}

The centrality dependence of event-by-event particle ratio fluctuations 
at 158$A$ GeV ($\sqrt{s_{NN}}$ = 17.3 GeV) is presented in Fig. 
\ref{ratio3}. The absolute values of fluctuations rise towards 
peripheral collisions, as in UrQMD. The same multiplicity scaling (as in 
Figs. \ref{ratio1}, \ref{ratio2}) seems to hold for all three particle 
ratio fluctuations (black, solid lines in Fig. \ref{ratio3}). This is 
compatible with the hypothesis that at constant energy the underlying 
correlations are not significantly changing with the system 
size.

\subsection{Average $p_T$ and multiplicity fluctuations}
\label{phiomega}

At the CP enlarged fluctuations of multiplicity and mean transverse 
momentum are expected \cite{SRS}. The NA49 experiment used the 
scaled variance of multiplicity distributions $\omega$ and the $\Phi_{p_{T}}$ 
measure to quantify multiplicity and average $p_T$ fluctuations, 
respectively (see \cite{kg_qm09} for details).  
The position of the chemical freeze-out point in the ($T - \mu_B$) 
diagram can be varied by changing the energy and the size of the 
colliding system \cite{beccatini, kg_qm09}. 
$T_{chem}$ decreases from p+p to Pb+Pb interactions at top SPS energy 
and $\mu_B$ decreases with increasing energy in Pb+Pb collisions. 
Therefore NA49 analyzed the energy ($\mu_B$) dependence of 
$\omega$ and $\Phi_{p_T}$ for central Pb+Pb collisions, and their 
system size ($T_{chem}$) dependence (p+p, central C+C, Si+Si, and Pb+Pb) 
at the highest SPS energy.

There are no indications of the CP in the energy dependence of 
multiplicity and mean $p_T$ fluctuations in central Pb+Pb collisions.
However, the system size dependence of both measures at 158$A$ GeV 
shows a maximum for C+C and Si+Si interactions \cite{kg_qm09}. 
The peak is even two times higher for all charged than for negatively 
charged particles \cite{kg_qm09} as expected for the CP 
\cite{SRS}. This result is consistent with a CP location near the 
freeze-out point of p+p interactions at the top SPS energy ($T=$178 
MeV, $\mu_B=250$ MeV) (the theoretical magnitude of the CP effect has a 
maximum close to Si+Si instead of p+p system due to the fact that the 
correlation length in the model monotonically decreases with decreasing 
size of the colliding system (see \cite{kg_qm09} for details).

\subsection{Azimuthal angle fluctuations}

The main motivation of studying azimuthal event-by-event fluctuations 
was to search for plasma instabilities \cite{instab}, critical point and 
onset of deconfinement, and flow fluctuations \cite{ffluct}. NA49 
evaluated the $\Phi$ measure of fluctuations (instead of 
using $p_T$, as in section \ref{phiomega}, one uses azimuthal angle $\phi$). 
There are several background effects that can influence the $\Phi_{\phi}$ 
measure, among them resonance decays, flow, (di-)jets, momentum 
conservation, quantum statistics. All of them were studied in 
\cite{phi_models}. 

\begin{figure}[ht]
\centering
\vspace{-0.3cm}
\includegraphics[width=0.45\textwidth]{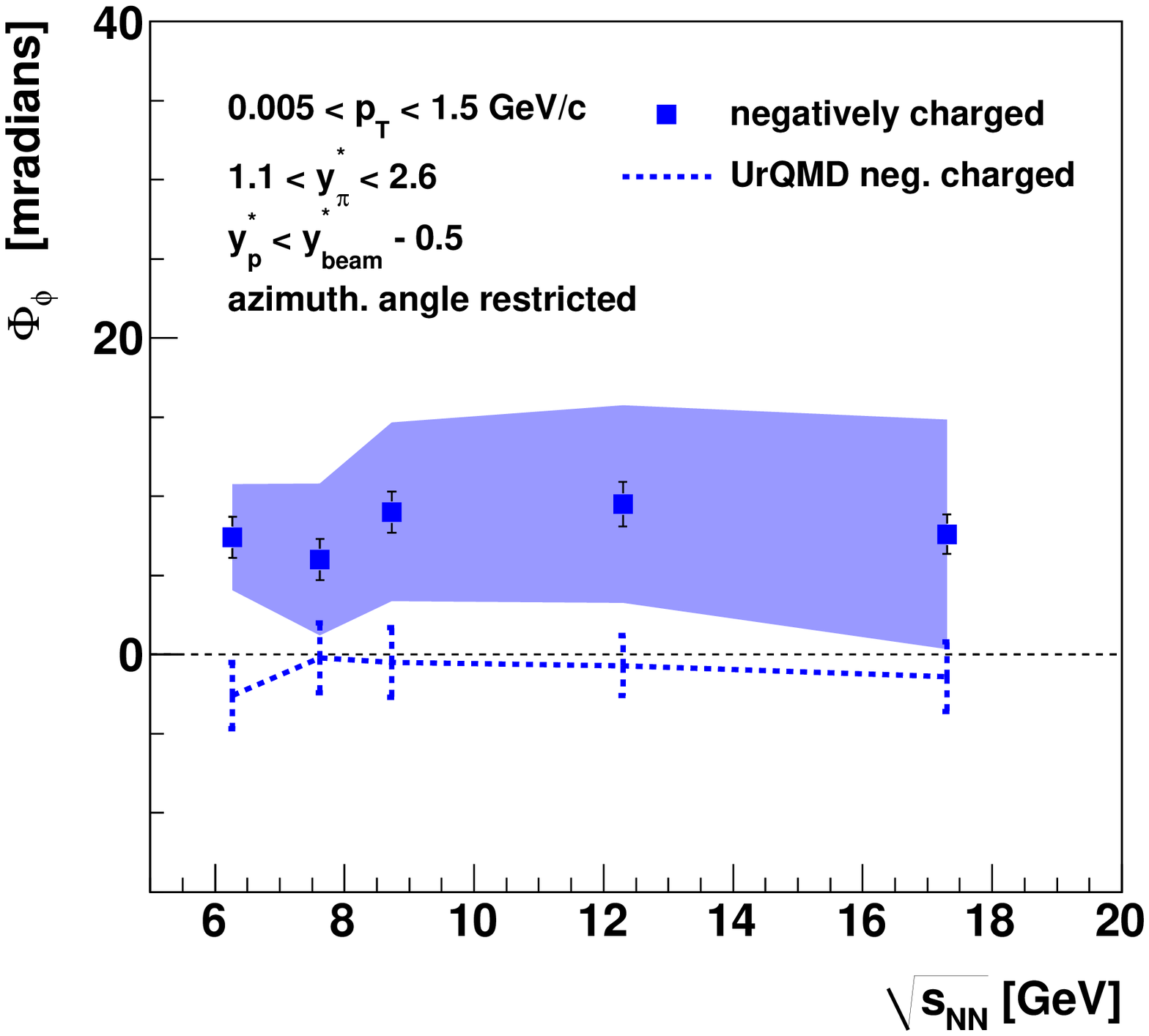}
\includegraphics[width=0.45\textwidth]{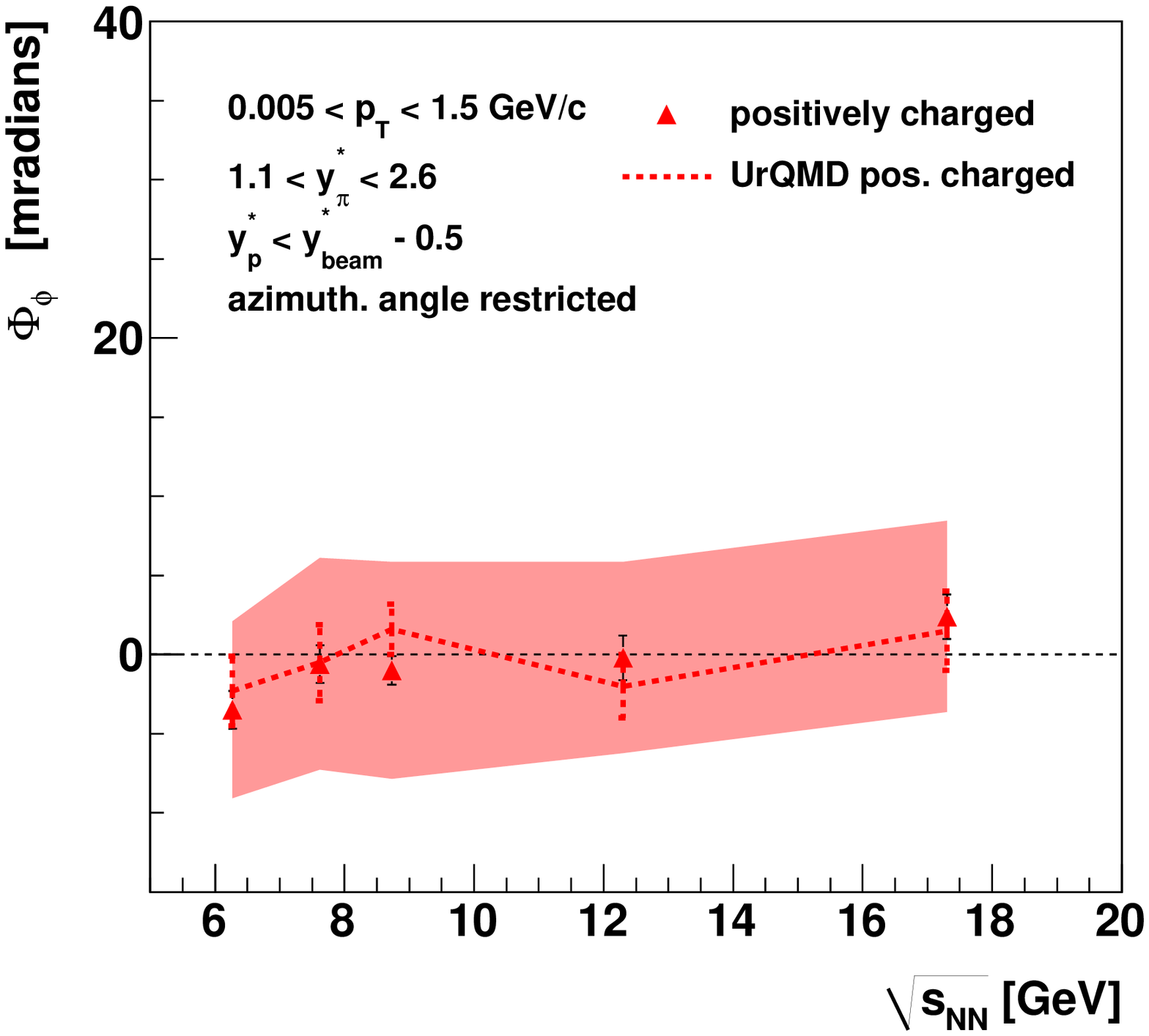}
\vspace{-0.7cm}
\caption[]{\footnotesize Energy dependence of azimuthal fluctuations. 
Forward rapidity, limited azimuthal acceptance (as in 
\cite{fluct_energy}). The same acceptance for data and UrQMD.}
\label{fifi_energy}
\end{figure}

Figure \ref{fifi_energy} shows the energy dependence of $\Phi_{\phi}$ 
for the 7.2\% most central Pb+Pb interactions. Color bands represent 
systematic errors. The values for positive particles are consistent 
with zero but for negative particles $\Phi_{\phi}$ is positive. No 
collision energy dependence of the fluctuations is observed. 

\begin{figure}[ht]
\centering
\vspace{-0.3cm}
\includegraphics[width=0.45\textwidth]{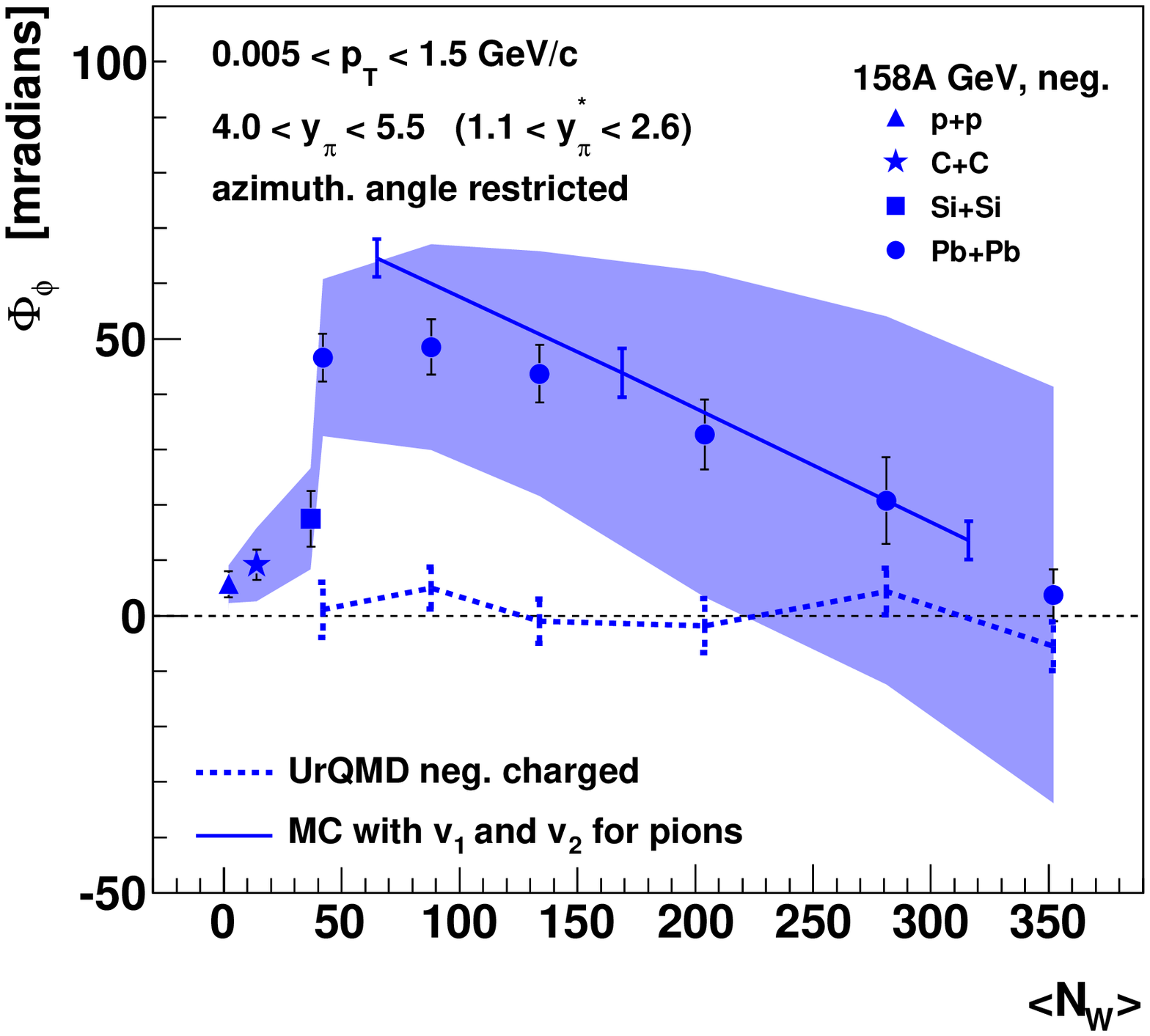}
\includegraphics[width=0.45\textwidth]{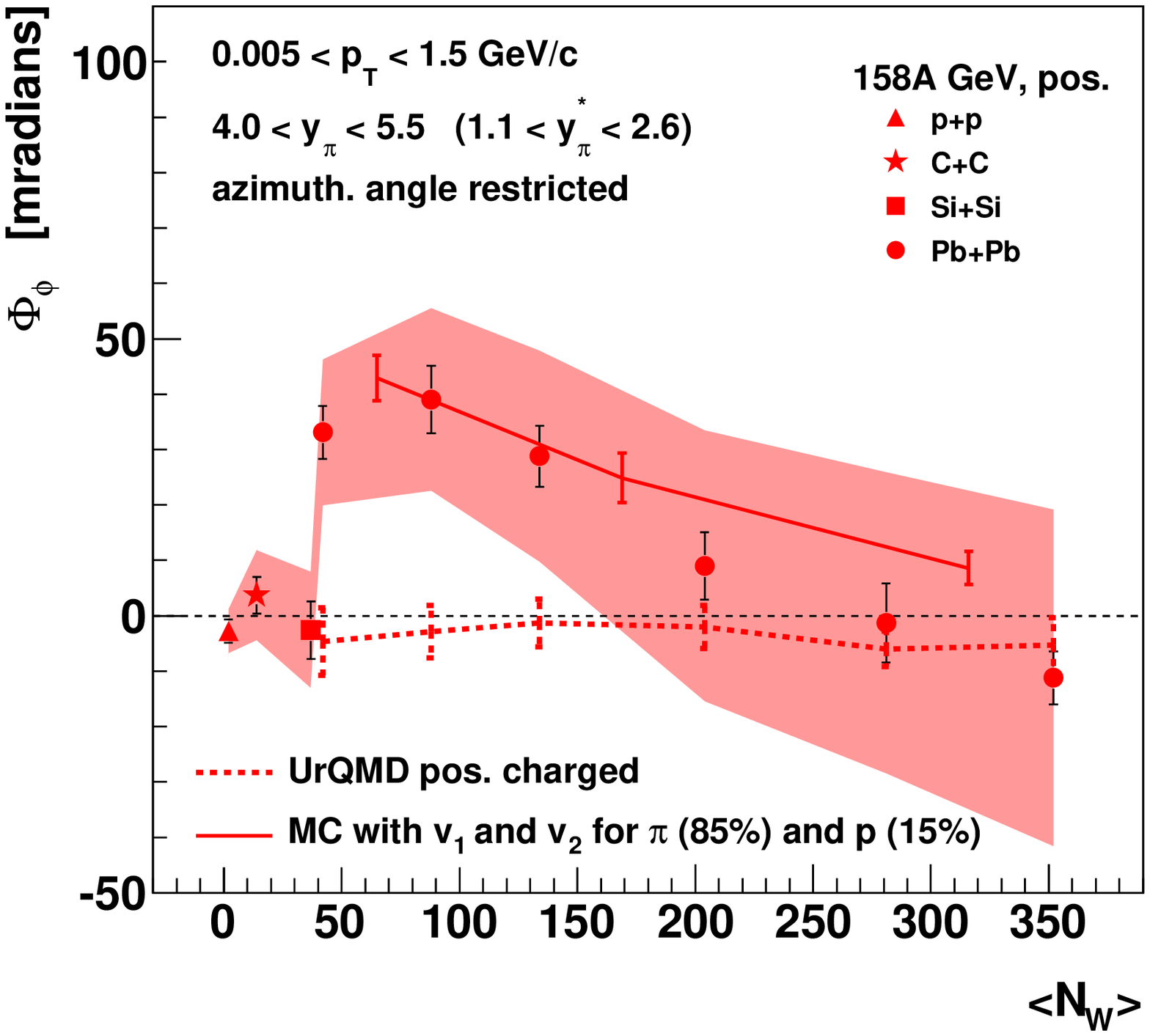}
\vspace{-0.7cm}
\caption[]{\footnotesize System size dependence of azimuthal 
fluctuations. Forward rapidity, limited azimuthal acceptance (as in 
\cite{fluct_size}). The same acceptance for data and UrQMD.}
\label{fifi_size}
\end{figure}

The system size and centrality dependence of $\Phi_{\phi}$ at the top 
SPS energy is presented in Fig. \ref{fifi_size}. 
For Pb+Pb collisions, the sample of events was split into 
six centrality classes. Figure \ref{fifi_size} shows positive 
$\Phi_{\phi}$ values with a maximum for peripheral Pb+Pb interactions. 
The data are not explained by the UrQMD model. However, the magnitude 
of $\Phi_{\phi}$ is reproduced by the effect of directed and elliptic flow 
\cite{wroclaw}. The difference between positive and negative particles 
is also reproduced and it is caused by a 15\% admixture of protons 
among positive particles (in the MC model calculation~\cite{wroclaw} $v_1$ and $v_2$ 
values for pions and protons at forward rapidity were taken from 
\cite{alexander}).

\subsection{Pion-pion intermittency signal}

It was suggested that the analog of critical opalescence 
may be detectable through intermittency analysis in $p_T$ space. 
Significant $\sigma$-field fluctuations are expected at the CP 
(density fluctuations of zero mass $\sigma$-particles 
produced in abundance at the CP) \cite{inter_th}. $\sigma$ particles at 
$T<T_c$ may reach the two-pion threshold ($2m_\pi$) and then decay into 
two pions, therefore density fluctuations of di-pions with 
$m_{\pi^{+}\pi^{-}}$ close to the two pion mass incorporate $\sigma$--field 
fluctuations at the CP. Local density fluctuations are expected 
both in configuration and momentum space.

\begin{wrapfigure}{r}{6.cm}
\vspace{-1cm}
\includegraphics[scale=0.55]{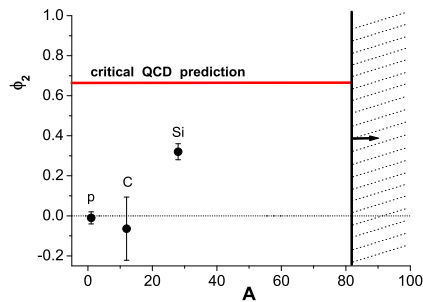}
\vspace{-1cm}
\caption[]{\footnotesize Intermittency signal in p+p, and 10\% most 
central C+C and Si+Si interactions at 158$A$ GeV \cite{inter_ex}.}
\label{inter}
\vspace{-0.3cm}
\end{wrapfigure}

The NA49 experiment searched for an intermittency signal in transverse 
momentum space of reconstructed di-pions ($\pi^{+}\pi^{-}$ pairs) with 
invariant mass just above $2m_\pi$ \cite{inter_ex}. The analysis was 
performed for p+p, C+C and Si+Si interactions at 158$A$ GeV. First, for 
each event all possible pairs with $m_{\pi^{+}\pi^{-}}$ in a small 
kinematic window above two-pion threshold were selected. Then second 
factorial moments $F_2 (M)$ in transverse momentum space were computed 
for real data and for artificially produced mixed events where only 
statistical fluctuations are present. The combinatorial background 
subtracted (by use of mixed events) moments $\Delta F_{2}$ in 
transverse momentum space are expected to follow a power-law behavior 
$\Delta F_{2 } \sim (M^2)^{\phi_2}$, with $\phi_2=2/3$ for systems 
freezing-out at CP \cite{inter_th}.

Figure \ref{inter} shows that $\Delta F_{2}$ for Si+Si at the top SPS 
energy measures fluctuations approaching in size the prediction of 
critical QCD (the remaining departure, $\phi_{2, max} \approx  0.33 \pm 
0.04$ instead of $2/3$, may be due to freezing out at a distance from 
the CP). As expected, the analysis of Si+Si events generated 
via the HIJING model shows
no intermittency signal
($\phi_{2} \approx 0.02 \pm 0.09$). NA49 (net)proton intermittency 
analysis is in progress.

\section{Summary}

The NA49 discovery of the energy threshold for deconfinement is now 
confirmed. The results from the RHIC Beam Energy Scan agree with NA49 
measurements on the onset of deconfinement. LHC data confirm 
the interpretation of the structures observed at low SPS energies as 
due to onset of deconfinement. 

New NA49 results on fluctuations were presented. The energy and the 
system size dependence of $K/\pi$ and $p/\pi$ fluctuations can be 
described in a simple multiplicity scaling model. In contrast, 
$K/p$ fluctuations show a deviation from this scaling and change sign 
close to the onset of deconfinement energy; is the underlying 
correlation physics changing with energy? 
For central $A+A$ collisions fluctuations of average $p_T$, 
multiplicity, and multiplicity of low mass $\pi^{+}\pi^{-}$ pairs tend 
to a maximum in Si+Si collisions at 158$A$ GeV. Thus the critical point 
may be accessible at SPS energies. This result is a strong 
motivation for future experiments and in fact, the NA49 
efforts will be continued by the ion program of the NA61/SHINE 
experiment \cite{MG_QM2011}.

\vspace{0.2cm}
{\footnotesize {\bf Acknowledgments:} This work was partially supported 
by Polish Ministry of Science and Higher Education under grant N 
N202 204638.}

\end{document}